\documentclass{PoS}

\usepackage{amssymb,amsmath,graphicx,epsfig,bm,color}
\usepackage[caption=false]{subfig}

\renewcommand{\d}{\mathrm{d}}
\newcommand{\be}{\begin{equation}}
\newcommand{\ee}{\end{equation}}
\newcommand{\bea}{\begin{eqnarray}}
\newcommand{\eea}{\end{eqnarray}}

\newcommand{\pup}{p^\uparrow}

\def\lsim{\mathrel{\rlap{\lower4pt\hbox{\hskip1pt$\sim$}}\raise1pt\hbox{$<$}}}
\def\gsim{\mathrel{\rlap{\lower4pt\hbox{\hskip1pt$\sim$}}\raise1pt\hbox{$>$}}}
\def\nostrocostruttino#1\over#2{\mathrel{\mathop{\kern 0pt \rlap
{\hbox{$#1$}}} \hbox{\kern-.135em $#2$}}}

\title{The gluon Sivers function and its process dependence from RHIC data}

\ShortTitle{The gluon Sivers function and its process dependence from RHIC data}

\author{\speaker{Cristian Pisano}\\
        Dipartimento di Fisica, Universit\`a di Cagliari and INFN, Sezione di Cagliari\\
        Cittadella Universitaria, I-09042 Monserrato (CA), Italy\\
        E-mail: \email{cristian.pisano@ca.infn.it}}

\author{Umberto D'Alesio\\
        Dipartimento di Fisica, Universit\`a di Cagliari and INFN, Sezione di Cagliari\\
        Cittadella Universitaria, I-09042 Monserrato (CA), Italy\\
        E-mail: \email{umberto.dalesio@ca.infn.it}}
        
\author{Carlo Flore\\
        Dipartimento di Fisica, Universit\`a di Cagliari and INFN, Sezione di Cagliari\\
        Cittadella Universitaria, I-09042 Monserrato (CA), Italy\\
        E-mail: \email{carlo.flore@ca.infn.it}}        

\author{Francesco Murgia\\
        INFN, Sezione di Cagliari,
        Cittadella Universitaria, I-09042 Monserrato (CA), Italy\\
        E-mail: \email{francesco.murgia@ca.infn.it}} 
        
 \author{Pieter Taels\\
        INFN, Sezione di Cagliari,
        Cittadella Universitaria, I-09042 Monserrato (CA), Italy\\
        E-mail: \email{pieter.taels@ca.infn.it}}

\abstract{We present a phenomenological analysis of available data on single spin asymmetries for pion and $D$ meson production in proton-proton collisions at RHIC within the color gauge invariant generalized parton model (CGI-GPM), which includes initial (ISIs) and final (FSIs) state interactions in a transverse momentum dependent formalism.  This study allows us, for the  first time, to put preliminary constraints on the two independent gluon Sivers functions entering the model. We also present a comparison with the simpler generalized parton model (without ISIs and FSIs).}

\FullConference{XXVII International Workshop on Deep-Inelastic Scattering and Related Subjects - DIS2019\\
		8-12 April, 2019\\
		Torino, Italy}

\begin{document}

\section{Introduction and formalism}

Transverse single-spin asymmetries (SSAs) in high-energy lepton-proton and proton-proton collisions are useful tools to probe the three-dimensional structure of the proton.  In this contribution we will focus on the inclusive production of pions and $D$ mesons in proton-proton scattering, {\it i.e.}\ $p^\uparrow p \to h\, X$ with $h= \pi$ or $D$. For these processes, SSAs are defined as
\begin{equation}
A_N \equiv \frac{\d\sigma^\uparrow-\d\sigma^\downarrow}{\d\sigma^\uparrow+\d\sigma^\downarrow} \equiv\, \frac{ \d\Delta\sigma}{ 2 \d\sigma}\,,
\end{equation}
with $\d\sigma^{\uparrow (\downarrow)}$ being the cross section in which one of the incoming protons is transversely polarized with respect to the production plane along a direction $\uparrow$ ($\downarrow$).  

For the calculation of the asymmetries, we adopt the generalized parton model (GPM)~\cite{DAlesio:2007bjf}, which assumes the validity of factorization of soft and hard parts, and includes both spin and transverse momentum effects. If the transversely polarized proton moves along the $\hat z$-axis and we denote by $\hat f_{a/\pup}\,(x_a, \bm k_{\perp a})$  the number density of its parton $a$, with light-cone momentum fraction $x_a$ and transverse momentum $\bm k_{\perp a} = k_{\perp a} (\cos\phi_a, \sin\phi_a)$,  the numerator of $A_N$ is sensitive to 
\bea
\Delta \hat f_{a/\pup}\,(x_a, \bm k_{\perp a}) &\equiv&
\hat f_{a/\pup}\,(x_a, \bm k_{\perp a}) - \hat f_{a/p^\downarrow}\,
(x_a, \bm k_{\perp a})
=  -2 \, \frac{k_{\perp a}}{M_p} \, f_{1T}^{\perp a} (x_a, k_{\perp a}) \>
\cos\phi_a \, ,
\label{defsiv}
\eea
where $M_p$ is the proton mass and $f_{1T}^{\perp a} (x_a, k_{\perp a})$ is  the Sivers function~\cite{Sivers:1989cc}, which  describes the azimuthal distribution of an unpolarized parton $a$ inside a transversely polarized proton. The Sivers function fulfills the following po\-si\-tivity bound
\begin{equation}
2\, \frac{k_{\perp a}}{M_p}\, \vert f_{1T}^{\perp a} (x_a, k_{\perp a})\vert \le   \hat f_{a/\pup}\,(x_a, \bm k_{\perp a}) \,+\,\hat f_{a/p^\downarrow} \, (x_a, \bm k_{\perp a})  \, \equiv 2\,  f_{a/p}\,(x_a, \bm k_{\perp a})\,,
\label{eq:posbound}
\end{equation}
with $f_{a/p}\,(x_a, \bm k_{\perp a})$ being the transverse momentum dependent distribution (TMD) of an unpolarized parton $a$ inside an unpolarized proton.

In the original formulation of the GPM, the Sivers function is taken to be universal. On the other hand, we expect that universality could  be broken by the presence of nonperturbative initial (ISIs) and final (FSIs) state interactions of the active partons in a specific process with the polarized proton remnants.  
The effects of ISIs and FSIs  have been initially taken into account in  the calculation of the {\em quark} Sivers function~\cite{Gamberg:2010tj,D'Alesio:2011mc}, by using a one-gluon exchange approximation. In this approach,  known as color gauge invariant generalized parton model (CGI-GPM),  the quark Sivers function is still universal, but is convoluted with modified partonic cross sections  into which the process dependence is absorbed. 

This model has been also extended to the gluon sector~\cite{DAlesio:2017rzj,DAlesio:2018rnv}.  Since for three colored gluons there are two different ways of forming a color singlet state, either through an $f$-type (totally antisymmetric, even under charge conjugation) or a  $d$-type  (symmetric and odd under $C$-parity) color combination, one needs  to introduce two independent gluon Sivers functions: $f_{1T}^{\perp g\, (f)}$~and~$f_{1T}^{\perp g\, (d)}$.

\section{Single spin asymmetries within the CGI-GPM framework}
In this section, we provide the leading order (LO) expressions for the numerators of the SSAs for the process $p^\uparrow p \to h\, X$, which can be used to extract information on the so-far poorly known gluon Sivers function.  Within the GPM one can write, schematically,
\begin{align}
 \d\Delta\sigma \, \propto\, \sum_{a,b,c,d}\, \left  (- \frac{k_{\perp\,a}}{M_p}\right )  f_{1T}^{\perp\,a}(x_a,  k_{\perp a}) \cos\phi_a\otimes 
 f_{b/p}(x_b, \bm k_{\perp b}) \otimes H_{ab\to c d }^U \otimes D_{h/c}(z,\bm k_{\perp h})\,,
\label{eq:num-SSA}
\end{align}
where  $D_{h/c}$ is the  fragmentation function of parton $c$ into $h$,  which depends on the variables $z$ (the  light-cone momentum fraction  of $c$ carried by $h$)  and  $\bm k_{\perp h}$ (the transverse momentum of $h$ with respect to the direction of $c$). Moreover,  the symbol $\otimes$ stands for a convolution in the transverse momenta and light-cone momentum fractions, while the hard functions $H^U_{ab\to c d }$ are related to the unpolarized cross sections for the partonic processes $ab\to cd$.

If the parton $a$ inside the polarized proton is a quark or an antiquark,  the numerator of the asymmetry in the CGI-GPM  can be formally obtained from Eq.~(\ref{eq:num-SSA}) by means of the substitution
\begin{align}
f_{1T}^{\perp\,q} \otimes H_{qb\to c d}^U   \longrightarrow f_{1T}^{\perp\,q}\otimes H_{qb\to cd}^{{\rm Inc}}\,,
\end{align}
where $H_{qb\to cd}^{{\rm Inc}}$ is the modified hard function for the process ${qb\to cd}$. On the other hand, if  $a=g$, one has to consider  two contributions,   
\begin{align}
f_{1T}^{\perp\,g} \otimes H_{gb\to c d}^U   \longrightarrow f_{1T}^{\perp\,g\,(f)}\otimes H_{gb\to cd}^{{\rm Inc} \,(f)}\, \, + \,\, f_{1T}^{\perp\,g\,(d)} \otimes H_{gb\to cd}^{{\rm Inc}\,(d)},
\end{align}
corresponding to the $f$- and $d$-type gluon Sivers functions.

Concerning $p^\uparrow p\to \pi\, X$, the LO $f$-type hard functions for the gluon induced subprocesses can be written in terms of the usual Mandelstam variables as follows~\cite{DAlesio:2018rnv}
\begin{align}
&H_{gq\to gq}^{\text{Inc} \,(f)} = H_{g\bar{q}\to g\bar{q}}^{\text{Inc} \,(f)}  =\! -\frac{\hat s^2+\hat u^2}{4 \hat s \hat u} \bigg (  \frac{\hat s^2}{\hat t^2} + \frac{1}{N_c^2} \bigg ) \,, \qquad \qquad\,\, H_{gq\to qg}^{\text{Inc} \,(f)} = H_{g\bar{q}\to \bar{q} g}^{\text{Inc} \,(f)}  = -\frac{\hat s^4-\hat t^4}{4 \hat s \hat t\hat u^2} \,,  \nonumber\\
&H_{gg\to q \bar q}^{\text{Inc} \,(f)}  = H_{gg\to \bar{q} q}^{\text{Inc} \,(f)}=\!-\frac{N_c}{4 (N_c^2-1)} \frac{\hat t ^2+ \hat u^2}{ \hat t \hat u} \bigg (\frac{\hat t^2}{\hat s^2}\, +\frac{1}{ N_c^2} \bigg ), \,
H_{gg\to gg}^{\text{Inc} \,(f)} = \frac{N_c^2}{N_c^2-1}\bigg (\frac{\hat t}{\hat u} - \frac{\hat s}{\hat u}\bigg) \frac{(\hat s^2 +  \hat s \hat t + \hat t^2)^2}{\hat s^2 \hat t^2}\, ,
\end{align}
where $N_c$ is the number of colors. For the $d$-type hard functions, we have 
\begin{align}
&H_{gq\to gq}^{\text{Inc} \,(d)} = \!-H_{g\bar{q}\to g\bar{q}}^{\text{Inc} \,(d)}= \frac{\hat s^2 + \hat u^2}{ 4 \hat s \hat u} \! \bigg ( \frac{ \hat s^2-2\hat u^2}{\hat t^2}\! +\! \frac{1}{N_c^2}  \bigg ), \,\,\, \,
 H_{gq\to q g}^{\text{Inc} \,(d)} = \!-H_{g\bar{q}\to \bar{q} g}^{\text{Inc} \,(d)} = \!-\frac{\hat s^2 + \hat t^2}{4 \hat s \hat t }\!\!  \left (\frac{\hat s^2 \!+ \!\hat t^2}{\hat u^2}  \!- \!\frac{2}{N_c^2}  \right )\!,\nonumber\\
& H_{gg\to q \bar q}^{\text{Inc} \,(d)}  =\! -H_{gg\to \bar{q} q}^{\text{Inc} \,(d)} =\!-\frac{N_c}{4 (N_c^2-1)}\frac{\hat t ^2+ \hat u^2}{ \hat t \hat u} \bigg ( \frac{\hat t^2-2 \hat u^2}{\hat s^2} \!+ \!\frac{1}{N_c^2} \bigg ), \,\, \, \,H_{gg\to gg}^{\text{Inc} \,(d)} = 0\,, \label{eq:Hgggg}
\end{align}
showing that, in principle, both gluon Sivers distributions can contribute to the asymmetry. Analogous quantities for the quark induced subprocesses have been included in our calculation. They are not shown here, but can be found in Ref.~\cite{Gamberg:2010tj}. Moreover, we point out that, although  $A_N$ can also originate from the fragmentation of transversely polarized quarks through the competing Collins mechanism~\cite{Collins:1992kk}, this effect is negligible for the RHIC kinematics under study here~\cite{DAlesio:2018rnv}. 

Turning to $p^\uparrow\, p \,\to D\,X$,  at LO the underlying subprocesses are $q\bar q \to c\bar c$ and $gg\to c \bar c$. The modified hard functions for the dominant gluon fusion processes read 
\begin{align}
H^{{\rm Inc} \,(f)}_{gg\to c\bar c} & = H^{{\rm Inc}\,(f)}_{gg\to \bar c c} = -\frac{N_c}{4(N_c^2-1)}\,  \frac{1}{\tilde t \tilde u} \, \left (  \frac{\tilde t^2}{\tilde s^2}+ \frac{1}{N_c^2}\right ) \left ( \tilde t^2 + \tilde u^2 + 4 m_c^2 \tilde s - \frac{4 m_c^4 \tilde s^2}{\tilde t \tilde u }    \right )\, ,\nonumber \\
H^{{\rm Inc} \,(d)}_{gg\to c\bar c} & =- H^{{\rm Inc}\,(d)}_{gg\to \bar c c} = -\frac{N_c}{4(N_c^2-1)}\,  \frac{1}{\tilde t \tilde u} \, \left (  \frac{\tilde t^2 - 2 \tilde u^2}{\tilde s^2}+ \frac{1}{N_c^2}\right ) \left ( \tilde t^2 + \tilde u^2 + 4 m_c^2 \tilde s - \frac{4 m_c^4 \tilde s^2}{\tilde t \tilde u }    \right )\,,
\label{eq:Hfd}
\end{align}
where we have introduced the invariants 
\begin{equation}
\tilde s  \equiv (p_a+p_b)^2 = \hat s\,,\qquad   \tilde{t} \equiv (p_a-p_c)^2-m_c^2= \hat t -m_c^2\,,\qquad \tilde{u} \equiv  (p_b-p_c)^2-m_c^2 = \hat u -m_c^2\,,
\end{equation}
and $m_c$ is the charm mass. 
Notice that all the contributions to $A_N$, other than the Sivers one, are strongly suppressed after integration over transverse momenta and can therefore be neglected~\cite{DAlesio:2017rzj}.

\begin{figure}[t]
\begin{center}
\includegraphics[trim = 1.cm 0.5cm 1cm 0.5cm, clip, width=7cm]{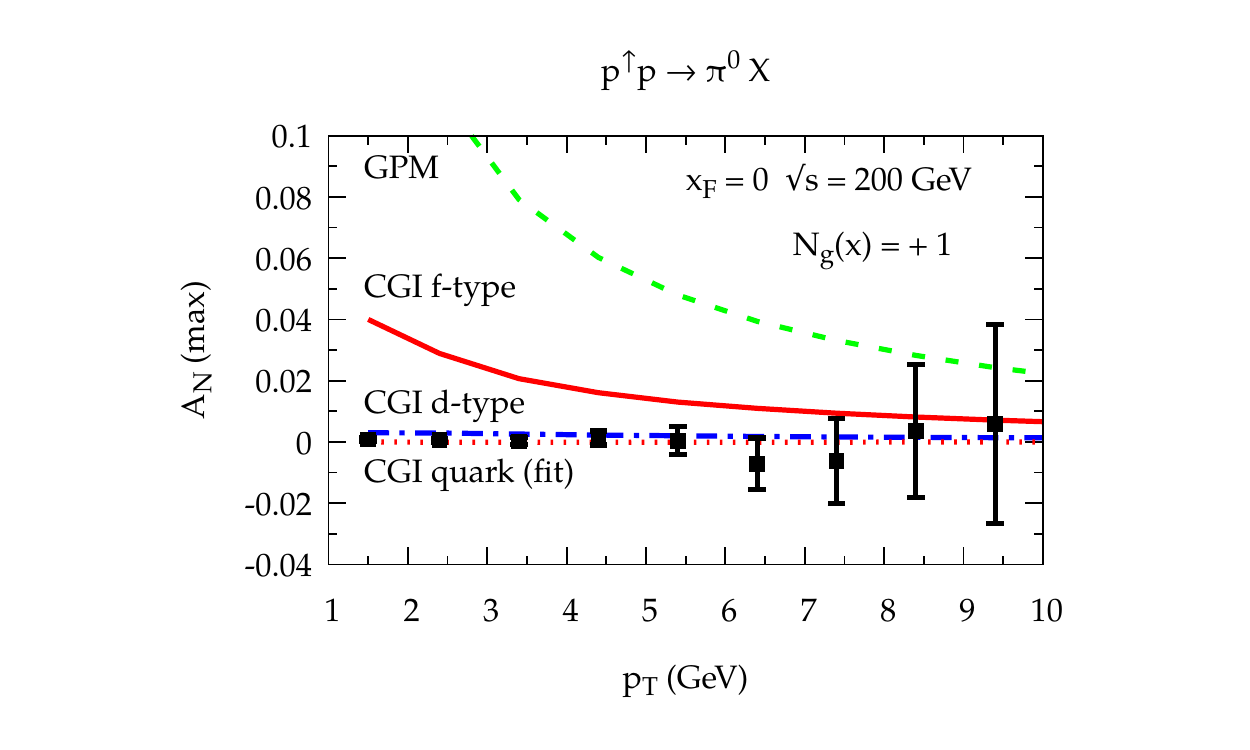}
\includegraphics[trim = 1.cm 0.5cm 1cm 0.5cm, clip, width=7cm]{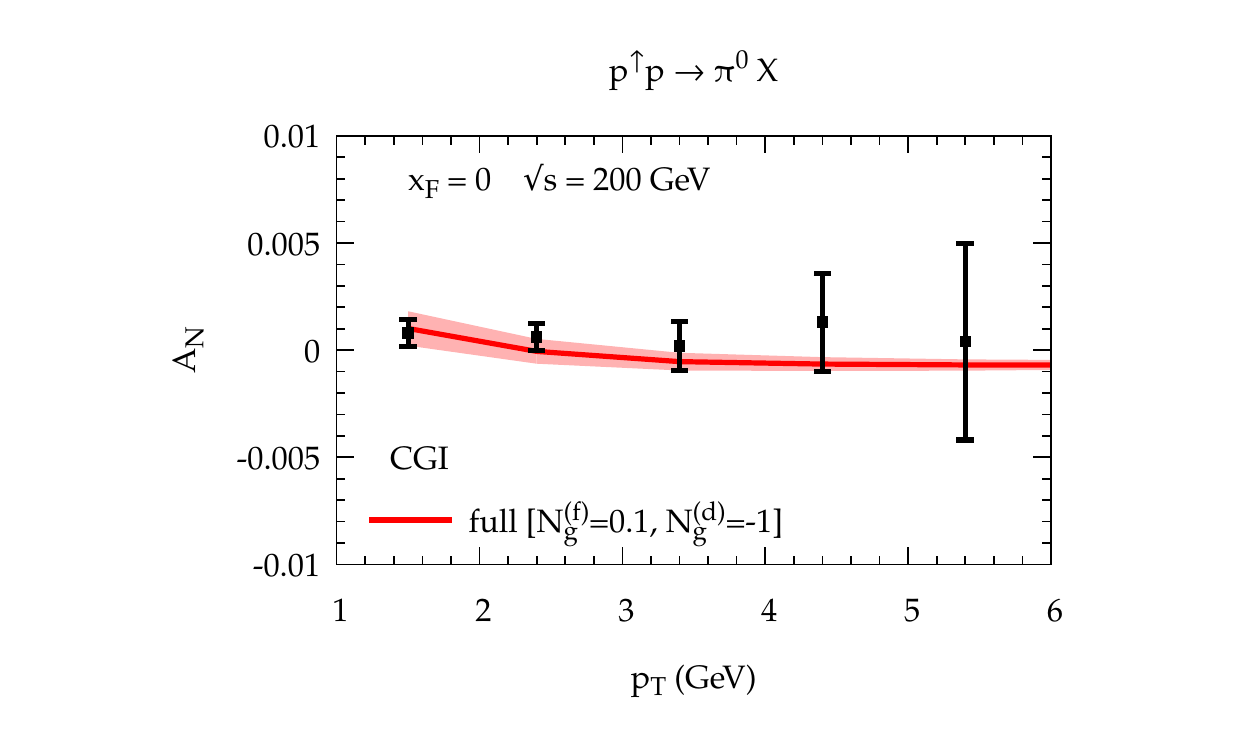}
\caption{Maximized gluon Sivers contributions to $A_N$ for 
$p^\uparrow p\to \pi^0\, X$ as a function of $p_T$,  together with the  quark Sivers effect  (left panel);  results for $A_N$ obtained with  a reduced $f$-type and a negative saturated  $d$-type gluon Sivers functions, {\it i.e.}\ with ${\cal N}_g^{(f)} =0.1$ and  ${\cal N}_g^{(d)} =-1$ (right panel). The shaded area describes  a $\pm 20$\% uncertainty on ${\cal N}_g^{(f)}$. Data are from Ref.~\cite{Adare:2013ekj}.}
\label{fig:AN-pi}
\end{center}
\end{figure}

\section{Numerical results}
In this section, we show how a combined analysis of SSAs for midrapidity pion and $D$-meson production can constrain the two gluon Sivers functions in the CGI-GPM. We assume that both distributions can be parametrized in term 
of the unpolarized collinear gluon density $f_{g/p}(x)$ as 
\begin{equation}
\Delta^N\! f_{g/p^\uparrow}(x,k_\perp) \equiv   
\left (-2\frac{k_\perp}{M_p}  \right )f_{1T}^{\perp\,g} (x,k_\perp)  = 2 \, {\cal N}_g(x)\,f_{g/p}(x)\,
h(k_\perp) \,\frac{e^{-k_\perp^2/\langle k_\perp^2 \rangle_g}}
{\pi \langle k_\perp^2 \rangle_g}\,,
\label{eq:siv-par-1}
\end{equation}
with
\begin{equation}
{\cal N}_g(x) = N_g x^{\alpha}(1-x)^{\beta}\,
\frac{(\alpha+\beta)^{(\alpha+\beta)}}
{\alpha^{\alpha}\beta^{\beta}}\;\;\;\;\;\; h(k_\perp) = \sqrt{2e}\,\frac{k_\perp}{M'}\,e^{-k_\perp^2/M'^2}\,,
\label{eq:nq-coll}
\end{equation}
where $|N_g|\leq 1$. We also define the additional parameter $ \rho = M'^2/(\langle k_\perp^2 \rangle_g +M'^2)$, with $0 < \rho < 1$.

We first consider the PHENIX data for the process $p^\uparrow p\to \pi^0\, X$ at $\sqrt s=200$ GeV and central rapidities as a function of $p_T$ ~\cite{Adare:2013ekj}. In order to assess the role of the  $f$- and $d$-type Sivers functions, in the left panel of Fig.~\ref{fig:AN-pi}
we maximize their effects  by saturating the positivity bound 
for their $x$-dependent parts (i.e.~${\cal N}_g(x)=\pm 1$), assuming $\rho = 2/3$. It turns out that the $d$-type contribution to $A_N$ is suppressed because of  the cancellation between the $gq$ and $g\bar q$ channels, entering with a relative opposite sign,
and the absence of the $gg\to gg$ contribution, relevant at moderate values of $p_T$, as evident from~\eqref{eq:Hgggg}. 
The quark Sivers effect, determined  from fits to semi-inclusive deep inelastic scattering data, is completely negligible. In contrast, the GPM framework gives the largest maximized effect, 
since its partonic contributions are all positive and unsuppressed. SSAs compatible with the very tiny and \emph{positive} data points could be obtained,  
in a conservative scenario,  if a relative cancellation between the $f$- and $d$-type Sivers functions occurs, 
with a strongly reduced and positive $f$-type Sivers function.
The results for ${\cal N}_g^{(f)}(x)=+0.1$ and ${\cal N}_g^{(d)}(x)=-1$ are shown in the right panel of Fig.~\ref{fig:AN-pi}, 
together with an overall uncertainty band of about $\pm$20\% on ${\cal N}_g^{(f)}$.

\begin{figure}[t]
\begin{center}
\includegraphics[trim = 1.cm 0.5cm 1cm 0.5cm, clip, width=7cm]{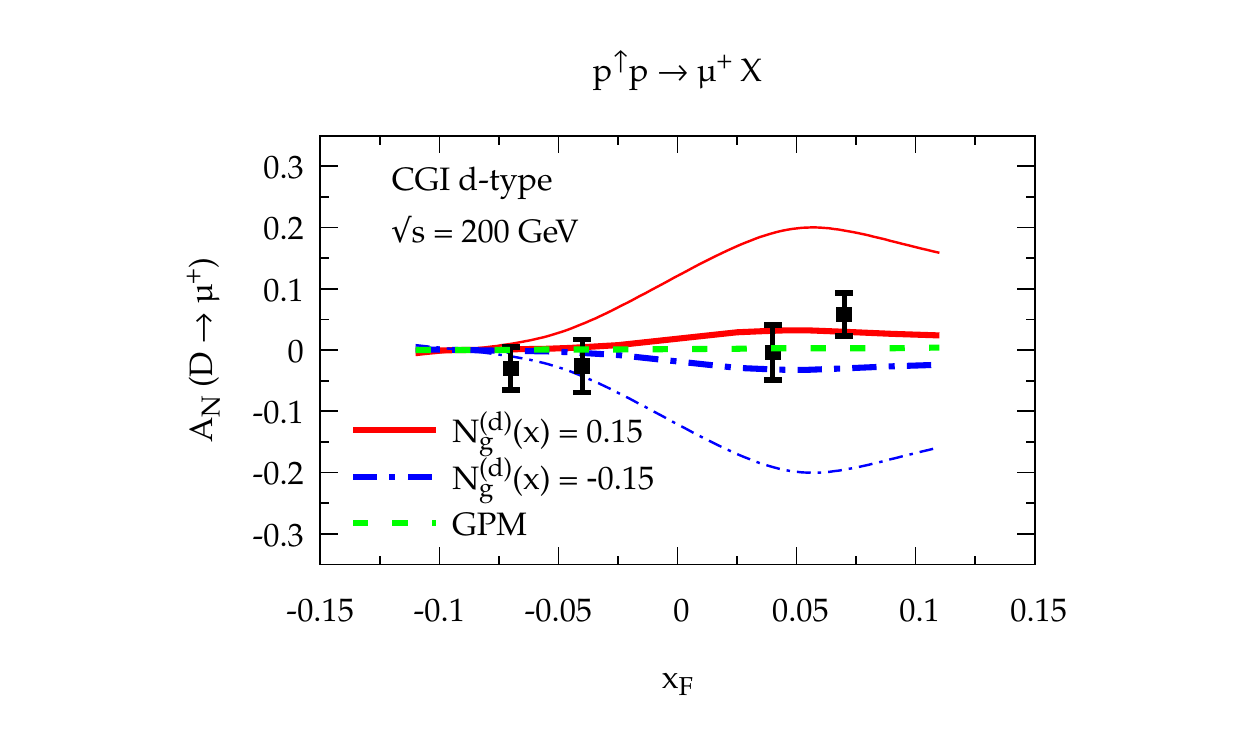}
\includegraphics[trim = 1.cm 0.5cm 1cm 0.5cm, clip, width=7cm]{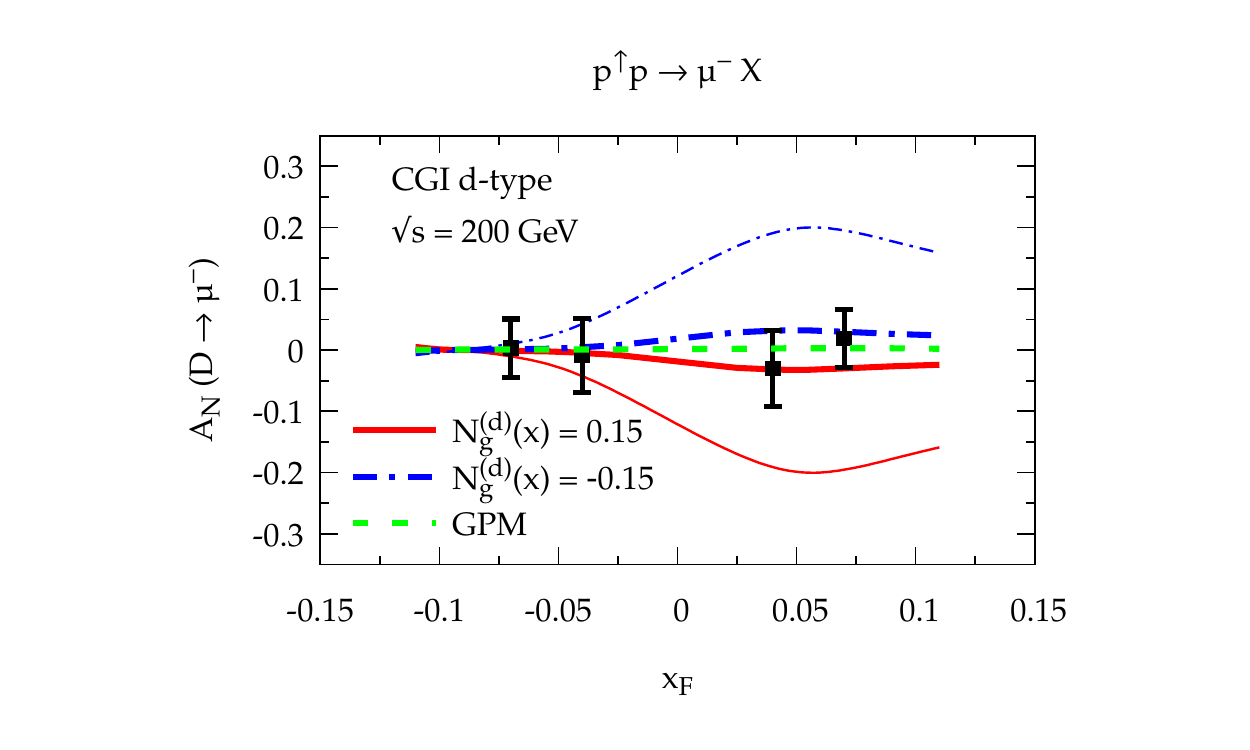}
\caption{Contributions of $f_{1T}^{\perp g\, (d)}$ to $A_N$ for $p^\uparrow p\to \mu^+\, X$ (left panel) and 
$p^\uparrow p\to \mu^-\, X$ (right panel) from $D$-meson decay as a function of $x_F$. The maximized effects correspond to ${\cal N}_g^{(d)} =+1$ (thin solid lines) and  ${\cal N}_g^{(d)}=-1$ (thin dash-dotted lines).
Predictions within the GPM are also presented. Data are from Ref.~\cite{Aidala:2017pum}.}
\label{fig:AN-muon}
\end{center}
\end{figure}

Our results for $D$-meson production have to be converted to SSAs for their muon decay products for a direct comparison with the data~\cite{Aidala:2017pum}.  The number and precision of the available data points do not allow for a true fit. However, 
important information can be obtained by adopting a very simple scenario with ${\cal N}_g^{(f,d)}(x)\equiv N_g^{(f,d)}$.
In Fig.~\ref{fig:AN-muon} the maximized $d$-type contributions are shown by thin lines:  
 the full one corresponds to ${\cal N}_g^{(d)} =+1$,  while the dash-dotted  one to ${\cal N}_g^{(d)} = -1$.  
We note that the data at positive values of the Feynman variable $x_F$ are largely overestimated.
On the other hand, the choice $|{\cal N}_g^{(d)}| \leq 0.15$, 
with a slight preference for positive values because of the positive $\mu^+$ data point, allows for a fair agreement with the data, as shown by the thick lines  in Fig.~\ref{fig:AN-muon} for both $\mu^+$ (left panel) and $\mu^-$ (right panel) production. The $f$-type contribution with ${\cal N}_g^{(f)} = 0.1$ is suppressed and therefore not shown explicitly. Moreover,  predictions within the GPM approach are also presented, showing a good agreement with the data.  We note that the GPM gluon Sivers function is  also parametrized according to  \eqref{eq:siv-par-1}-\eqref{eq:nq-coll},  with $\langle k_\perp^2\rangle_g = 1$ GeV$^2$ and $N_g=0.25$,  $\alpha = 0.6$,  $\beta = 0.6$,  $\rho = 0.1$,  as extracted in Ref.~\cite{DAlesio:2018rnv} from the neutral pion data~\cite{Adare:2013ekj}. 

Going back to the pion SSAs, we find that, by varying ${\cal N}_g^{(d)}$ 
in the range $-0.15\le {\cal N}_g^{(d)}  \le +0.15$ with $\rho=2/3$, a very good description of both 
the $\mu^\pm$ and $\pi^0$ data is possible if ${\cal N}_g^{(f)}$ 
varies in the range $+0.05 \ge {\cal N}_g^{(f)}\ge -0.01$. More precisely, 
\be
\label{eq:par_gsf_CGI}
{\cal N}_g^{(d)} = - 0.15\,  \Longrightarrow  {\cal N}_g^{(f)} = +0.05\,, \hspace*{1cm}
{\cal N}_g^{(d)} = + 0.15\,  \Longrightarrow  {\cal N}_g^{(f)} = -0.01\,.
\ee
Hence we find that a stronger reduction of $f_{1T}^{\perp g\, (f)}$ emerges from this combined analysis.

\begin{figure}[t]
\begin{center}
\includegraphics[trim = 1.cm 0.15cm 1cm 0.15cm, clip, width=8cm]{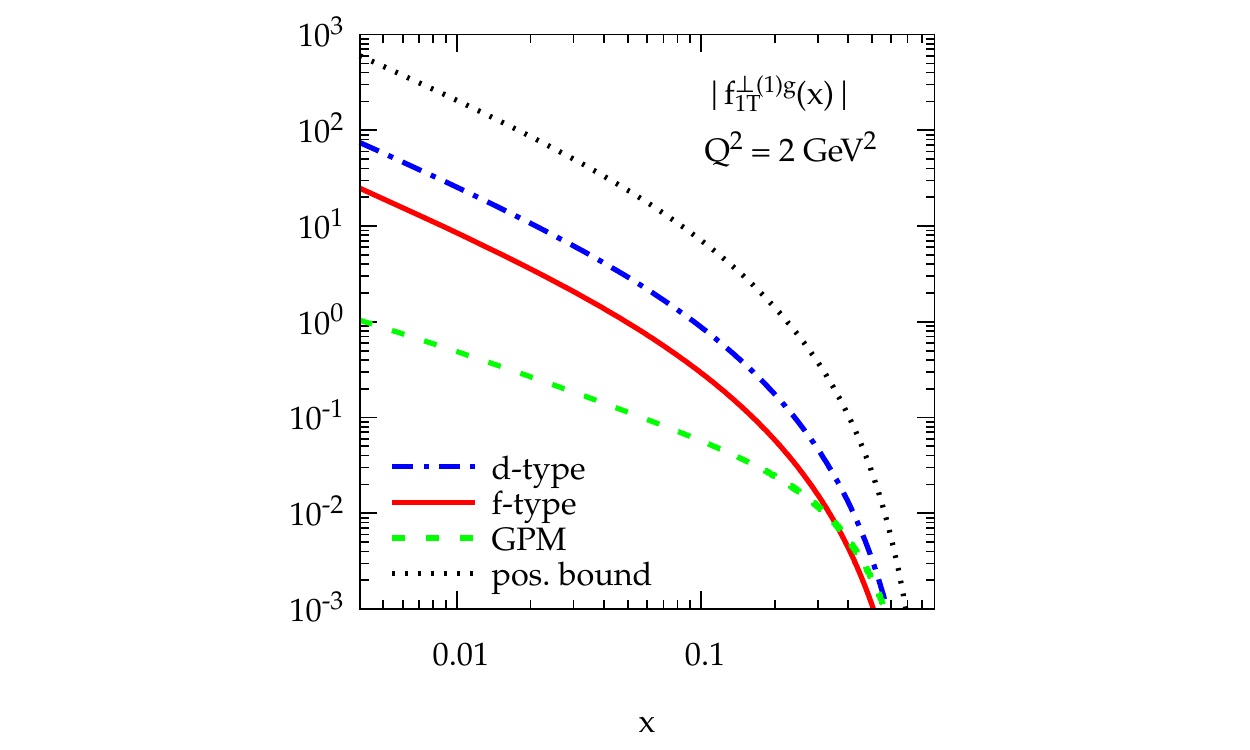}
\caption{Values of the first transverse moments of the resulting gluon Sivers functions at $Q^2 = 2$ GeV$^2$.
}
\label{fig:1stmom}
\end{center}
\end{figure}

The absolute values of the first transverse moments of  the Sivers functions obtained above,  
\be
\Delta^N \! f_{g/p^\uparrow}^{(1)}(x) = 
\int \d^2 \bm{k}_\perp \frac{k_\perp}{4 M_p} \Delta^N \! f_{g/\pup}(x,k_\perp) \equiv - f_{1T}^{\perp (1) g}(x) \, .
\label{siversm1}
\ee
are presented in  Fig.~\ref{fig:1stmom} for the GPM 
and the CGI-GPM frameworks, with $N_g^{(f)}=0.05$ and $N_g^{(f)}=0.15$, 
together with the positivity bound in \eqref{eq:posbound}.  Notice that the gluon Sivers distribution in the GPM,  for which the hard partonic functions are all positive, is the smallest. 

To conclude, we have performed a  first attempt  towards a quantitative study of the process dependence of the gluon Sivers  function. 
Although we are not able yet to clearly discriminate between the GPM and the CGI-GPM frameworks, our results are encouraging; they have been used to provide predictions for $A_N$ in $p^\uparrow p\to J/\psi\,X$ and $p^\uparrow p \to \gamma\, X$~\cite{DAlesio:2018rnv}, both under study  at RHIC.

\begin{thebibliography}{99}

\bibitem{DAlesio:2007bjf}
  U.~D'Alesio and F.~Murgia,
  \emph{Azimuthal and Single Spin Asymmetries in Hard Scattering Processes},
  \emph{Prog.\ Part.\ Nucl.\ Phys.}\  {\bf 61} (2008) 394
  [{\tt arXiv:0712.4328 [hep-ph]}].

\bibitem{Sivers:1989cc}
  D.~W.~Sivers,
  \emph{Single Spin Production Asymmetries from the Hard Scattering of Point-Like Constituents},
  \emph{Phys.\ Rev.\ D} {\bf 41} (1990) 83.

\bibitem{Gamberg:2010tj}
  L.~Gamberg and Z.~B.~Kang,
  \emph{Process dependent Sivers function and implication for single spin asymmetry in inclusive hadron production},
  \emph{Phys.\ Lett.\ B} {\bf 696} (2011) 109
 [{\tt arXiv:1009.1936 [hep-ph]}].

\bibitem{D'Alesio:2011mc}
  U.~D'Alesio, L.~Gamberg, Z.~B.~Kang, F.~Murgia and C.~Pisano,
  \emph{Testing the process dependence of the Sivers function via hadron distributions inside a jet},
  \emph{Phys.\ Lett.\ B} {\bf 704} (2011) 637
  [{\tt arXiv:1108.0827 [hep-ph]}].

\bibitem{DAlesio:2017rzj}
  U.~D'Alesio, F.~Murgia, C.~Pisano and P.~Taels,
  \emph{Probing the gluon Sivers function in $p^\uparrow p\to J/\psi\,X$ and $p^\uparrow p \to D\,X$},
  \emph{Phys.\ Rev.\ D} {\bf 96} (2017) 036011
  [{\tt arXiv:1705.04169 [hep-ph]}].

\bibitem{DAlesio:2018rnv}
  U.~D'Alesio, C.~Flore, F.~Murgia, C.~Pisano and P.~Taels,
  \emph{Unraveling the Gluon Sivers Function in Hadronic Collisions at RHIC}, \emph{Phys.\ Rev.\ D} {\bf 99} (2019) 036013 
 [{\tt arXiv:1811.02970 [hep-ph]}].
   
\bibitem{Collins:1992kk}
  J.~C.~Collins,
  \emph{Fragmentation of transversely polarized quarks probed in transverse momentum distributions},
  \emph{Nucl.\ Phys.\ B} {\bf 396} (1993) 161
  [{\tt hep-ph/9208213}].
 
\bibitem{Adare:2013ekj}
{\scshape PHENIX} Collaboration, \emph{{Measurement of transverse-single-spin
  asymmetries for midrapidity and forward-rapidity production of hadrons in
  polarized p+p collisions at $\sqrt{s}=$200 and 62.4 GeV}},
 {\emph{Phys. Rev.\ D} {\bf 90} (2014) 012006}
  [{\tt arXiv:1312.1995 [hep-ex]}].

\bibitem{Aidala:2017pum}
 {\scshape PHENIX} Collaboration, 
  \emph{Cross section and transverse single-spin asymmetry of muons from open heavy-flavor decays in polarized $p$+$p$ collisions at $\sqrt{s}=200$ GeV},
  \emph{Phys.\ Rev.\ D} {\bf 95} (2017) 112001
  [{\tt arXiv:1703.09333 [hep-ex]}].

\end{thebibliography}
\end{document}